\documentclass[12pt]{article} 

\usepackage{hyperref,natbib}
\usepackage{graphicx}
\usepackage{amsmath, wrapfig,amssymb,multirow}
\usepackage[margin=1in]{geometry}

\usepackage{chngcntr}
\counterwithout{figure}{section}

\newcommand{\instructions}[1]{}

\usepackage[parfill]{parskip}
\usepackage{epsfig, subfigure}
\usepackage{amsfonts}

\usepackage{url}
\usepackage{authblk}

\usepackage{booktabs}
\usepackage{bigstrut}
\usepackage{multirow}
\usepackage{threeparttable} 
\usepackage[format=hang,labelfont=bf]{caption}
\usepackage{colortbl}
\definecolor{light-gray}{gray}{0.75}
\usepackage{url}
\usepackage{float}
\floatstyle{boxed}
\usepackage{arydshln}
\newfloat{algbox}{tbp}{lop}

\begin{document}
\pagenumbering{roman}

\title{{\bf {\sc InSilicoVA:}}\\[5pt] \Large{A Method to Automate Cause of Death Assignment \\for Verbal Autopsy \vspace{2cm}}}

\author[1,4,5,6,7,*]{Samuel J. Clark}
\author[1,2]{Tyler McCormick}
\author[2]{Zehang Li}
\author[2,3]{\authorcr Jon Wakefield}

\affil[1]{Department of Sociology, University of Washington}
\affil[2]{Department of Statistics, University of Washington}
\affil[3]{Department of Biostatistics, University of Washington}
\affil[4]{Institute of Behavioral  Science (IBS), University of Colorado at Boulder}
\affil[5]{MRC/Wits Rural Public Health and Health Transitions Research Unit (Agincourt), \authorcr School of Public Health, Faculty of Health Sciences, University of the Witwatersrand}
\affil[6]{ALPHA Network, London School of Hygience and Tropical Medicine, London, UK}
\affil[7]{INDEPTH Network, Accra, Ghana}
\affil[*]{Correspondence to: \texttt{work@samclark.net}}

\date{\vspace{2cm} August 24, 2013}

\maketitle


\newpage

\centerline{{\sc Abstract}}
Verbal autopsies (VA) are widely used to provide cause-specific mortality estimates in developing world settings where vital registration does not function well.  VAs assign cause(s) to a death by using information describing the events leading up to the death, provided by care givers.  Typically physicians read VA interviews and assign causes using their expert knowledge.  Physician coding is often slow, and individual physicians bring bias to the coding process that results in non-comparable cause assignments.  These problems significantly limit the utility of physician-coded VAs.  A solution to both is to use an algorithmic approach that formalizes the cause-assignment process.  This ensures that assigned causes are comparable and requires many fewer person-hours so that cause assignment can be conducted quickly without disrupting the normal work of physicians. Peter Byass' InterVA method \citep{byass2012strengthening} is the most widely used algorithmic approach to VA coding and is aligned with the WHO 2012 standard VA questionnaire  \citep{leitao2013whoStandard}.  

The statistical model underpinning InterVA can be improved; uncertainty needs to be quantified, and the link between the population-level CSMFs and the individual-level cause assignments needs to be statistically rigorous.  Addressing these theoretical concerns provides an opportunity to create new software using modern languages that can run on multiple platforms and will be widely shared.  Building on the overall framework pioneered by InterVA, our work creates a statistical model for automated VA cause assignment.  

\vfill

\footnotesize
\centerline{\sc{Acknowledgments}}
Preparation of this manuscript was partially supported by the Bill and Melinda Gates Foundation.  The authors are grateful to Peter Byass, Basia Zaba, Kathleen Kahn, Stephen Tollman, Adrian Raftery, Philip Setel and Osman Sankoh for helpful discussions.
\normalsize

\newpage

\tableofcontents

\newpage

\pagenumbering{arabic}

\section{Introduction}

Verbal autopsy (VA) is a common approach for determining cause of death in regions where deaths are not recorded routinely. VAs are a standardized questionnaire administered to caregivers, family members, or others knowledgeable of the circumstances of a recent death with the goal of using these data to infer the likely causes of death \citep{byass2012strengthening}.  The VA survey instrument asks questions related to the deceased individual's medical condition (did the person have diarrhea, for example) and related to other factors surrounding the death (did the person die in an automotive accident, for example).  VA has been widely used by researchers in Health and Demographic Surveillance Sites (HDSS), such as the INDEPTH Network \citep{Sankoh2012}, and has recently received renewed attention from the World Health Organization through the release of an update to the widely used standard VA questionnaire (see \url{http://www.who.int/healthinfo/statistics/verbalautopsystandards/}).  The main statistical challenge with VA data is to ascertain patterns in responses that correspond to a pre-defined set of causes of death.  Typically the nature of such patterns is not known \emph{a priori} and measurements are subject to various types of measurement error, which are discussed in further detail below.

There are two credible methods for automating the assignment of cause of death from VA data.  InterVA~\citep[see for example:][]{byass2003probabilistic,byass2006refining,byass2012strengthening} is a proprietary algorithm developed and maintained by Peter Byass, and it has been used extensively by both the ALPHA \citep{maher2010translating} and INDEPTH networks of HDSS sites and a wide variety of other research groups.  At this time InterVA appears to be the {\em de facto} standard method.  The Institute for Health Metrics and Evaluation (IHME) have proposed a number of additional methods \citep[for example:][]{flax, james, murray}, some of which build on earlier work by King and Lu~\citep{king1, king2}.   Among their methods, the Simplified Symptom Pattern Recognition (SSPR) method (Murray et al., 2011) is most directly comparable to InterVA and appears to have the best chance of becoming widely used.  The SSPR and related methods require a so-called `gold standard' database, a database consisting of a large number of deaths where the cause has been certified by medical professionals and is considered reliable, and further, where the symptoms for those deaths are also verifiable by medical professionals.  Deaths recorded in a gold standard database are typically in-hospital deaths.  Given regional variation in the prevalence and familiarity of medical professionals with certain causes of death, region-specific gold standard databases are also necessary.  The majority of public health and epidemiology researchers and officials do not have access to such a gold standard database, motivating development of methods to infer cause of death using VA data without access to gold standard databases. 

Aside from the challenges related to obtaining useful gold standard training data, using VA data to infer cause assignment is also statistically challenging because there are multiple sources of variation and error present in VA data. We identify three sources of variation and propose a novel method called {\bf InSilicoVA} to address these sources of variation.  First, though responses on the VA instrument may be the respondentÕs recollection or best guess about what has happened, they are not necessarily accurate.  This type of error is present in all survey data but may be especially pronounced in VA data because respondents have varying levels of familiarity with the circumstances surrounding a death.  Also, the definition of an event of interest may be different for each respondent.  A question about diarrhea, for example, requires the respondent to be sufficiently involved in the decedent's medical care to know this information and to have a definition of diarrhea that maps relatively well to accepted clinical standards.  A second source of variability arises from individual variation in presentation of diseases.  Statistical methods must be sufficiently robust to overfitting to appreciate that two individuals with the same cause of death may have slightly different presentations, and thus, different VA reports.  Third, like InterVA, InSilicoVA will use physician elicited conditional probabilities.  These probabilities, representing the consensus estimate of the likelihood that a person will experience a given symptom if they died from a certain cause, are unknown.  Simulation and results with data from the Agincourt MRC/Wits Rural Public Health and Heath Transitions Research Unit (Agincourt) indicate that results obtained with conditional probabilities assumed fixed and known (as is done in both InterVA and the SSPR method) can underestimate uncertainty in population cause of death distributions.  We evaluate the sensitivity to these prior probabilities.  

InSilicoVA incorporates uncertainty from these sources and propagates the uncertainty between individual cause assignments and population distributions.  Accounting for these sources of error produces confidence intervals for both individual cause assignments and population distributions.  These confidence intervals reflect more realistically the complexity of cause assignment with VA data.  A unified framework for both individual cause and population distribution estimation also means that additional information about individual causes, such as physician coded VAs, can easily be used in InSilicoVA, even if physician coded records are only available for a subset of cases.  Further, we can exactly quantify the contribution of each VA questionnaire item in classifying a case into a specific cause, affording the possibility for `item reduction' by identifying which inputs are most useful for discriminating between causes. This feature could lead to a more efficient, streamlined survey mechanism and set of conditional probabilities for elicitation.  

In Section \ref{sec:interva} we describe InterVA and present open challenges which we address with InSilicoVA.  In Section~\ref{sec:insilico} we present the InSilicoVA model.  Section \ref{sec:methods} describes applications of both methods to simulated and real data, and Section~\ref{sec:results} provides the results.  We conclude with a very brief discussion in Section~\ref{sec:discuss}.

\section{InterVA}\label{sec:interva}

\subsection{Byass' Description of InterVA}\label{sec:byassDescribesInterva}

Below is Byass' summary of the InterVA method presented in \cite{byass2012strengthening}.
\footnotesize
\begin{quote}
BayesÕ theorem links the probability of an event
happening given a particular circumstance with the
unconditional probability of the same event and the conditional
probability of the circumstance given the event.
If the event of interest is a particular cause of death,
and the circumstance is part of the events leading to
death, then BayesÕ theorem can be applied in terms of
circumstances and causes of death.
Specifically, if there are a predetermined set of possible
causes of death $C_1 \dots C_m$ and another set of indicators
$I_1 \dots I_n$ representing various signs, symptoms and circumstances
leading to death, then BayesÕ general theorem
for any particular $C_i$ and $I_j$ can be stated as:
\begin{equation}
\label{eqn:byassBayes}
P(C_i | I_j) = \frac{P(I_j | C_i) \times P(C_i)} {P(I_j | C_i) \times P(C_i) + P(I_j | !C_i) \times P(!C_i)}
\end{equation}
where $P(!C_i)$ is $(1-P(C_i))$.
Over the whole set of causes of death $C_1 \dots C_m$ a set
of probabilities for each $C_i$ can be calculated using
a normalising assumption so that the total conditional
probability over all causes totals unity:
\begin{equation}
\label{eqn:byassBayesSum}
P(C_i | I_j) = \frac{P(I_j | C_i) \times P(C_i)} {\sum_{i=1}^m{P(C_i)}}
\end{equation}
Using an initial set of unconditional probabilities for
causes of death $C_1 \dots C_m$ (which can be thought of as
$P(C_i | I_0)$) and a matrix of conditional probabilities $P(I_j | C_i)$
for indicators $I_1 \dots I_n$ and causes $C_1 \dots C_m$, it is possible
to repeatedly apply the same calculation process for each
$I_1 \dots I_n$ that applies to a particular death:
\begin{equation}
\label{eqn:byassAlgorithm}
P(C_i | I_{1 \dots n}) = \frac{P(I_j | C_i) \times P(C_i | I_{0 \dots n-1})} {\sum_{i=1}^m{P(C_i | I_{0 \dots n-1})}}
\end{equation}
This process typically results in the probabilities of
most causes reducing, while a few likely causes are characterised
by their increasing probabilities as successive
indicators are processed.
\end{quote}
\normalsize

In the same article Byass describes the process of defining the conditional probabilities $P(I_j | C_i)$.
\footnotesize
\begin{quote}
Apart from the mathematics, the major challenge in
building a probabilistic model covering all causes of
death to a reasonable level of detail lies in populating
the matrix of conditional probabilities $P(I_j | C_i)$. There is
no overall source of data available which systematically
quantifies probabilities of various signs, symptoms and
circumstances leading to death in terms of their associations
with particular causes. Therefore, these conditional
probabilities have to be estimated from a diversity of
incomplete sources (including previous InterVA models)
and modulated by expert opinion. In the various versions
of InterVA that have been developed, expert panels have
been convened to capture clinical expertise on the relationships
between indicators and causes. In this case,
an expert panel convened in
Geneva in December 2011 and continued to deliberate
subsequently, particularly considering issues that built
on previous InterVA versions. Experience has shown that
gradations in levels of perceived probabilities correspond
more to a logarithmic than linear scale, and in the
expert consultation for InterVA-4, we used a perceived
probability scale that was subsequently converted to
numbers on a logarithmic scale as shown below.
\captionsetup{font=footnotesize,margin=5.5cm,justification=raggedright}
\begin{table}[H]
\center
\footnotesize
\caption{InterVA Conditional Probability Letter-Value Correspondances}
\begin{tabular}{l l l}
\hline
Label & Value & Interpretation \\
\hline\hline
I & 1.0 & Always \\
A & 0.8 & Almost always \\
A & 0.5 & Common \\
A & 0.2 & \\
B & 0.1 & Often \\
B & 0.05 & \\
B & 0.02 & \\
C & 0.01 & Unusual \\
C & 0.005 & \\
C & 0.002 & \\
D & 0.001 &vRare \\
D & 0.0005 & \\
D & 0.0001 & \\
E & 0.00001 & Hardly ever \\
N & 0.0 & Never \\
\hline \hline
\end{tabular}
\label{tab:conditionalPs}
\end{table}
\captionsetup{font=normalsize}
\end{quote}
\normalsize

The physician-derived conditional probabilities that are supplied with the InterVA software \citep{2013interVA} are coded using the letter codes in the leftmost column of Table \ref{tab:conditionalPs}.

We rewrite, interpret and discuss the InterVA model below.

\subsection{Our Notation for InterVA:}

\begin{itemize}
\item Deaths: $y_j \ \ j \in \{1, \dots , J\}, \ \vec{Y} = [y_1, \dots , y_J]$
\item Signs/symptoms: $s_k \in \{0,1\}, \ k \in \{1, \dots , K\}, \ \vec{S} = [s_1, \dots , s_K]$
\item Causes: $c_n \ \ n \in \{1, \dots , N\}, \ \sum_{n=1}^{N}{c_{j,n}} = 1$
\item Fraction of all deaths that are cause $n$, the `cause-specific mortality fraction' (CSMF): $f_n \ \ n \in \{1, \dots , N\}, \ \vec{F} = [f_1, \dots , f_N], \ \sum_{n=1}^{N}{f_n} = 1$
\end{itemize}

\subsection{Our Description of InterVA Data Requirements:}\label{sec:intervaData}

\begin{enumerate}
\item For each death $y_j$, a VA interview produces a binary-valued vector of signs/symptoms: 
\begin{equation} \vec{S_j} = \{s_{j,1}, s_{j,2}, \dots s_{j,K}\} \label{eqn:intervaS} \end{equation}
$\mathbf{S}$ is the $J \times K$ matrix whose rows are the $\vec{S}_j$ for each death.  
\item A $K \times N$ matrix of conditional probabilities reflecting physicians' opinions about how likely a given sign/symptom is for a death resulting from a given cause:
\begin{equation}
\mathbf{P} = 
\begin{bmatrix}
\Pr(s_1 | c_1) & \Pr(s_1 | c_2) & \cdots & \Pr(s_1 | c_N) \\
\Pr(s_2 | c_1) & \Pr(s_2 | c_2) & \cdots & \Pr(s_2 | c_N) \\
\vdots & \vdots & \ddots & \vdots \\  
\Pr(s_K | c_1) & \Pr(s_K | c_2) & \cdots & \Pr(s_K | c_N)
\end{bmatrix}
\label{eqn:intervaP}
\end{equation}
As supplied with the InterVA software \citep{2013interVA} $\mathbf{P}$ does not contain internally consistent probabilities\footnote{ 
The $\mathbf{P}$ supplied with InterVA has many logical inconsistencies, for example situations where conditional probabilities should add up to equal another: $\Pr(\mbox{fast breathing for 2 weeks or longer} \ | \ \mbox{HIV}) + \Pr(\mbox{fast breathing for less than 2 weeks} \ | \ \mbox{HIV}) \neq \Pr(\mbox{fast breathing} \ | \ \mbox{HIV})$, or where they just do not make sense: $\Pr(\mbox{fast breathing for 2 weeks or longer} \ | \ \mbox{sepsis}) > \Pr(\mbox{fast breathing} \ | \ \mbox{sepsis}).$ The $\mathbf{P}$ supplied with InterVA-4 \citep{2013interVA} is a 254 $\times$ 69 matrix with 17,526 entries.  We have investigated automated ways of correcting the inconsistencies, but with every attempt we discover more, so we have concluded that the entries in $\mathbf{P}$ need to be re-elicited from physicians using an approach that ensures that they are consistent.}.  
This is easy to understand by noting that these probabilities are not derived from a well-defined event space that would constrain them to be consistent with one another.  As described by Byass above in Section \ref{sec:byassDescribesInterva} the physicians provide a `letter grade' for each conditional probability, and these correspond to a ranking of perceived likelihood of a given sign/symptom if the death is due to a given cause.  These letter grades are then turned into numbers in the range $[0,1]$ (NB: 0.0 and 1.0 are included) using Table \ref{tab:conditionalPs}. 

Consequently it is not possible to assume that the members of $\mathbf{P}$ will behave as expected when one attempts to calculate complements and use more than one in an expression in a way that {\em should} be consistent.
\item An initial guess of $\vec{F}$, $\vec{F}' = [f_n', \dots ,f_N']$
\end{enumerate}

\subsection{Our Presentation of the InterVA Model and Algorithm}

For a specific death $y_j$ we can {\em imagine} and examine the two-dimensional joint distribution $(c_{j,n},s_{j,k})$: 
\begin{eqnarray}
\Pr(c_{j,n} | s_{j,k}) & = & \frac{P(s_{j,k} | c_{j,n}) \cdot \Pr(c_{j,n})} {\Pr(s_{j,k})} \nonumber \\
& = & \frac{\Pr(s_{j,k} | c_{j,n}) \cdot \Pr(c_{j,n})} {\Pr(s_{j,k} | c_{j,n}) \cdot \Pr(c_{j,n}) + x \label{eqn:intervaPosterior}}
\end{eqnarray}
where 
\begin{equation}
x = \Pr(s_{j,k} | \neg c_{j,n}) \cdot \Pr(\neg c_{j,n})
\end{equation}

Looking on the RHS of (\ref{eqn:intervaPosterior}), we have $\Pr(s_{j,k} | c_{j,n})$ from the conditional probabilities from physicians and $\Pr(c_{j,n}) \approx f_n^{'}$.  If the conditional probabilities $\mathbf{P}$ were well-behaved, then 
\begin{equation}
x = \sum_{n'=1, \ n' \neq n}^{N}{\Pr(s_{j,k} | c_{j,n'}) \cdot \Pr(c_{j,n'})}
\end{equation}
However since the $\mathbf{P}$ supplied with the InterVA software \citep{2013interVA} are not consistent with one another this calculation does not produce useful results.

InterVA solves this with an arbitrary reformulation of the relationship.  For each death $y_j$ and over all signs/symptoms $s_k$ associated with $y_j$: 
\begin{equation}
\label{eqn:intervaStep1}
\forall \ (j,n): \mbox{Propensity}(c_{j,n} | \vec{S_j}) = f_n' \cdot \prod_{k=1}^K \left[ \Pr(s_{j,k} | c_{n}) \right]^{s_k}  \end{equation}%
For each death $y_j$ these `Propensities' do not add to 1.0 so they need to be normalized to produce well-behaved probabilities:
\begin{equation}
\label{eqn:IntervaStep2}
\forall \ (j,n): \ \Pr(c_{j,n}) = \frac {\mbox{Propensity}(c_{j,n} | \vec{S_j}) }{ \sum_{n=1}^N \mbox{Propensity}(c_{j,n} | \vec{S_j})}
\end{equation}
The population-level CSMFs $\vec{F}$ are calculated by adding up the results of calculating  (\ref{eqn:IntervaStep2}) for all causes for all deaths:
\begin{equation}
\label{eqn:intervaCsmf}
\forall \ n: \ f_n = \sum_{j=1}^J \Pr(c_{j,n})
\end{equation}

\subsection{Evaluation of InterVA Model \& Algorithm}

In effect what InterVA does is distribute a given death among a number of predefined causes.  The cause with the largest fraction is assumed to be the primary cause, followed with decreasing significance by the remaining causes in order from largest to smallest.  The conceptual construct of a `partial' death is central to InterVA and is interchanged freely with the {\em probability} of dying from a given cause.  This equivalence is not real and is at the heart of the theoretical problems with InterVA.

At a high level InterVA proposes a very useful solution to the fundamental challenge that all automated VA coding algorithms face - how to characterize and use the relationship that exists between signs/symptoms and causes of death.  In a perfect world we would have medically certified patient records that include the results of `real' autopsies, and we could use those to model this relationship and use the results of that model in our cause-assignment algorithms.  But in that perfect world there is no use for VA at all. So by definition we live in the world where that type of `gold standard' data do not and will not exist most of the time for most of the developing world where VAs are conducted.  Byass' solution to this is to accept the limitations on the expert knowledge that physicians can provide to substitute for gold standard data, and further, to elicit and then organize that information in a very useful format -- the conditional probabilities matrix $\mathbf{P}$ above in (\ref{eqn:intervaP}).  In sum, Byass has sorted through a variety of possible general strategies and settled on one that is both {\em doable} and produces useful results.

Where we contribute is to help refine the statistical and computational methods used to conduct the cause assignments.  In order to do that we have evaluated InterVA thoroughly, and we have identified a number of weaknesses that we feel can be addressed.  The brief list below is by necessity a blunt description of those weaknesses, which nevertheless do not reduce the importance of InterVA as described just above.

There are several theoretical problems with the InterVA model:
\begin{enumerate}
\item The derivation presented in (\ref{eqn:byassBayes}) through (\ref{eqn:byassAlgorithm}) is incorrect; in particular (\ref{eqn:byassBayesSum}) does not follow from (\ref{eqn:byassBayes}), and (\ref{eqn:byassAlgorithm}) does not follow from (\ref{eqn:byassBayesSum}).  As described just above, (\ref{eqn:intervaCsmf}) requires that probabilities be equated with fractional deaths, which is conceptually difficult. 
\item InterVA's statistical model is not `probabilistic' in the recognized sense because it does not include elements that can vary unpredictably, and hence there is no {\em randomness}.  Although $\mathbf{P}$ contains `probabilities' (see discussion with (\ref{eqn:intervaP}) above), these are not allowed to vary in the estimation procedure used to assign causes of death -- $\mathbf{P}$ is effectively a {\em fixed} input to the model.
\item Because the model does not contain features that are allowed to vary unpredictably, it is not possible to quantify uncertainty to produce probabilistic error bounds.
\item If we ignore the errors in the derivation and work with (\ref{eqn:intervaStep1}) - (\ref{eqn:intervaCsmf}) as if they were correct, there are additional problems. Equation (\ref{eqn:intervaStep1}) is at the core of InterVA and demonstrates two undesirable features:
\begin{enumerate}
\item For a specific individual the propensity for each cause is deterministically affected by $f_n'$, what Byass terms the `prior' probability of cause $n$, effectively a user-defined parameter of the algorithm.  This means that the final individual-level cause assignments are a deterministic transformation of the so-called `prior' -- i.e. the results are not only sensitive to but {\em depend directly} on the `prior'.
\item The expression in (\ref{eqn:intervaStep1}) captures only one valance in the relationship between signs/symptoms and a cause of death -- it  acknowledges and reacts only to the presence of a sign/symptom but not to its absence, effectively throwing away half of the information in the dataset.  To include information conveyed by the absence of a sign/symptom, (\ref{eqn:intervaStep1}) needs a term that involves something like `$1-\Pr(s_{j,k} | C_{n})$'. [The components of InSilicoVA in (\ref{eqn:insilicoIndividualProbCause}) and (\ref{eqn:insilicoGibbsF}) below include this term.] This is undesirable for two reasons: (1) signs/symptoms are not selecting causes that fit  {\em and} de-selecting causes that don't, but rather {\em just} de-selecting causes, and (2) the final probabilities are typically the product of a large number of very small numbers, and hence their numeric values can become extremely small, small enough to interact badly and unpredictable with the numerical storage/arithmetic capacity of the software and computers used to calculate them.  A simple log transformation would solve this problem. 
\end{enumerate}
\item Finally, (\ref{eqn:intervaCsmf}) is a deterministic transformation of the individual-level cause assignments to produce a population-level CSMF; simply another way of stating the individual-level cause assignments.  Because the individual-level cause assignments are not probabilistic, neither are the resulting CSMFs.
\end{enumerate}

In addition there are idiosyncrasies that affect the current implementation of InterVA \citep{2013interVA} and some oddities having to do with the matrix of conditional probabilities provided with Byass' InterVA.  We will not describe those here.

InSilicoVA is designed to overcome these problems and provide a valid statistical framework on which further refinements can be built.

\section{InSilicoVA}\label{sec:insilico}

InSilicoVA is a statistical model and computational algorithm to automate assignment of cause of death from data obtained by VA interviews.  Broadly the method aims to:
\begin{itemize}
\item Follow in the footsteps of InterVA building on its strengths and addressing its weaknesses.
\item Produce consistent, comparable cause assignments and CSMFs.
\item Be statistically and computationally valid and extensible.
\item Provide a means to quantify uncertainty.
\item Be able to function to assign causes to a single death.
\end{itemize}

The name `InSilicoVA' is inspired by `in-vitro' studies that mimic real biology but in more controlled circumstances (often on `glass' petri dishes). In this case we are assigning causes to deaths using a computer that performs the required calculations using a silicon chip.  Further, we owe a great debt to the InterVA (interpret VA) method that provides useful philosophical and practical frameworks on which to build the new method - so we have stuck to the structure of InterVA's name. 

\subsection{InSilicoVA Notation:}

\begin{itemize}
\item Deaths: $y_j \ \ j \in \{1, \dots , J\}, \ \vec{Y} = [y_1, \dots , y_J]$
\item Signs/symptoms: $s_k \in \{0,1\}, \ k \in \{1, \dots , K\}, \ \vec{S} = [s_1, \dots , s_K]$
\item Causes of death: $c_n \ \ n \in \{1, \dots , N\}, \ \vec{C} = [c_1 , \dots, c_N]$
\item For individual $j$, probability of cause $n$ given $\vec{S}_j$: $\ell_{j,n} = \Pr(y_j=c_n | \vec{S_j}), \ j \in \{1, \dots , J\}, \\ \ n \in \{1, \dots , N\}$, $\vec{L_j} = [l_{j,1}, \dots , l_{j,N}], \ \sum_{n=1}^{N}{\ell_{j,n}} = 1$ 
\item Count of all deaths that are cause $n$, the `cause-specific death count' (CSDC): \\ $m_n \ \ n \in \{1, \dots , N\}, \ \vec{M} = [m_1, \dots , m_N], \ \sum_{n=1}^{N}{m_n} = J$
\item Fraction of all deaths that are cause $n$, the `cause-specific mortality fraction' (CSMF): \\ $f_n \ \ n \in \{1, \dots , N\}, \ \vec{F} = [f_1, \dots , f_N], \ \sum_{n=1}^{N}{f_n} = 1$
\end{itemize}

\subsection{InSilicoVA Data:}

\begin{enumerate}
\item For each death $y_j$, the VA interview produces a binary-valued vector of signs/symptoms: 
\begin{equation}
\vec{S_j} = \{s_{j,1}, s_{j,2}, \dots s_{j,K}\} 
\label{eqn:insilicoS}
\end{equation}
$\mathbf{S}$ is the $J \times K$ matrix whose rows are the $\vec{S}_j$ for each death.  The columns of $\mathbf{S}$ are assumed to be independent given $\vec{C}$, i.e. there is no systematic relationship between the signs/symptoms for a given cause.  This is very obviously not a justifiable assumption.  Signs and symptoms come in characteristic sets depending on the cause of death, so there {\em is} some correlation between them, conditional on a given cause.  Nonetheless we assume independence in order to facilitate initial construction and testing of our model, and most pragmatically, so that we can utilize the matrix of conditional probabilities supplied by Byass with the InterVA software \citep{2013interVA} -- it is impossible to either regenerate or significantly improve upon these without significant resources with which to organize meetings of physicians with the relevant experience who can provide this information.  This $\vec{S}_j$ is the same as (\ref{eqn:intervaS}) used by InterVA.
\item A $K \times N$ matrix of conditional probabilities reflecting physicians' opinions about how likely a given sign/symptom is for a death resulting from a given cause:
\begin{equation}
\mathbf{P} = 
\begin{bmatrix}
\Pr(s_1 | c_1) & \Pr(s_1 | c_2) & \cdots & \Pr(s_1 | c_N) \\
\Pr(s_2 | c_1) & \Pr(s_2 | c_2) & \cdots & \Pr(s_2 | c_N) \\
\vdots & \vdots & \ddots & \vdots \\  
\Pr(s_K | c_1) & \Pr(s_K | c_2) & \cdots & \Pr(s_K | c_N)
\end{bmatrix}
\label{eqn:insilicoP}
\end{equation}
InSilicoVA assumes that the components of $\mathbf{P}$ are consistent with one another.  In the simulation study described in Section \ref{sec:simStudy}, we construct consistent values for $\mathbf{P}$, but when we test the model on real data in Section \ref{sec:aginStudy}, we have no  option other than using the inconsistent $\mathbf{P}$ supplied with the InterVA software \citep{2013interVA}.
\item An initial guess of $\vec{F}$, $\vec{F}' = [f_n', \dots ,f_N']$
\end{enumerate}

\subsection{InSilicoVA Algorithm}

We are interested in the joint distribution $(\vec{F},\vec{Y})$ given the set of observed signs/symptoms $\mathbf{S}$.  The posterior distribution is: 
\begin{eqnarray}
\Pr(\vec{F}, \vec{Y} | \mathbf{S}) &=& \frac{\Pr(\mathbf{S} | \vec{Y}, \vec{F}) \Pr(\vec{Y} | \vec{F}) \Pr(\vec{F})}{\Pr(\mathbf{S}) } \nonumber \\ 
& \propto& \Pr(\mathbf{S} | \vec{Y}, \vec{F}) \Pr(\vec{Y} | \vec{F}) \Pr(\vec{F}) \label{eqn:insilicoPosterior} \\
& = & \prod_{j=1}^J \Pr(\mathbf{S} | y_j,\vec{F}) \Pr(y_j | \vec{F}) \Pr(\vec{F})
\end{eqnarray}

Because individual cause assignments are independent, individual sign/symptom vectors $\vec{S_j}$ are independent from $\vec{F}$ (the CSMFs), and we have:
\begin{equation}
\label{eqn:insilicoSimplifiedPosterior}
\Pr(\vec{F}, \vec{Y} | \mathbf{S}) \propto \prod_{j=1}^J \Pr(\mathbf{S} | y_j) \Pr(y_j | \vec{F}) \Pr(\vec{F})
\end{equation}

We will use a Gibbs sampler to sample from this posterior as follows:
\begin{enumerate}
\item[G.1] start with an initial guess of $\vec{F}$, $\vec{F}'$
\item[G.2] sample $\vec{Y} | \vec{F}, \mathbf{S}$
\item[G.3] sample $\vec{F} | \vec{Y}, \mathbf{S}$
\item[G.4] repeat steps G.2 and G.3 until $\vec{F}$ and $\vec{Y}$ converge
\end{enumerate}

This algorithm is generic and allows a rich range of models.  For the moment the InsilicoVA model is:
\begin{eqnarray}
s_{j,k}|c_n  &\sim & \mbox{Bernoulli}(\Pr(s_{j,k} | c_n)) \label{eqn:symptomBernoulli} \\
y_j=c_n | \vec{F} &\sim & \mbox{Multinomial}_N(1,\vec{F} ) \label{eqn:causeAssignMultinomial} \\
\vec{F} &\sim & \mbox{Dirichlet}(\vec{\alpha}), \ \vec{\alpha} \ \mbox{is } N\mbox{-dimensional and constant} \label{eqn:insilicoCsfmAssignDirichlet}
\end{eqnarray} 

Then the posterior in (\ref{eqn:insilicoSimplifiedPosterior}) is:
\begin{equation}
\label{eqn:insilicoPosteriorComputation}
\Pr(\vec{F}, \vec{Y} | \mathbf{S}) \propto  \prod_{j=1}^J \prod_{k=1}^K \Pr(s_{j,k} | y_j=c_n) \Pr(y_j=c_n|\vec{F}) \Pr(\vec{F}) 
\end{equation}

This formulation is computationally efficient because of Multinomial/Dirichlet conjugacy, and because using Bayes rule we have, for step G.2:
\begin{equation} 
\label{eqn:insilicoDrawDeath} 
y_j=c_n | \vec{F},\vec{S}_j \sim \mbox{Multinomial}_N(1,\vec{L}_j) 
\end{equation}
where the $\ell_{j,n}$ that compose $\vec{L}_j$ are:
\begin{eqnarray}
\ell_{j,n} & = & \Pr(y_j=c_n | \vec{S_j}) \nonumber \\
& = & \frac {\Pr(y_j=c_n) \cdot \Pr(\vec{S_j} | y_j=c_n)} {\Pr(\vec{S_j})} \nonumber \\[6pt]
& & \mbox{substituting } f_n =\Pr(y_j=c_n) \mbox{ and using the data } \vec{S}_j \mbox{ and } \mathbf{P} \mbox{ to calculate} \nonumber \\
& & \mbox{the probability of a specific } \vec{S_j} \mbox{ given the cause assignment } y_j=c_n \nonumber \\[6pt]
& = & \frac{f_n \cdot  \prod_{k=1}^{K}{\left( \Pr(s_{j,k} | y_j=c_n)^{s_{j,k}} \cdot \left[1 - \Pr(s_{j,k} | y_j=c_n)\right]^{(1-s_{j,k})} \right)} }
{ \sum_{n'=1}^{N} f_{n'} \cdot \prod_{k=1}^{K}{\left( \Pr(s_{j,k} | y_j=c_{n'})^{s_{j,k}} \cdot \left[1 - \Pr(s_{j,k} | y_j=c_{n'})\right]^{(1-s_{j,k})} \right)}} \label{eqn:insilicoIndividualProbCause}
\end{eqnarray}

We can also derive from (\ref{eqn:insilicoPosteriorComputation}) the distribution of $\vec{F}$ conditional on $\vec{Y}$, for step G.3:
\begin{equation} 
\label{eqn:insilicoDrawCsmf} 
\vec{F} | \vec{Y}, \mathbf{S} \sim \mbox{Dirichlet}(\vec{M}+\vec{\alpha}) 
\end{equation}
where
\begin{equation}
\label{eqn:insilicoMn} 
m_n = \sum_{j=1}^{J}{[y_j=c_n]}, \mbox{ using Iverson's bracket notation:} [z] = \begin{cases} 1 \mbox{ if } z \mbox{ true} \\ 0 \mbox{ if } z \mbox{ false} \end{cases} 
\end{equation}

In summary, the Gibbs sampler proceeds given suitable initialization $\vec{F}'$ by:
\begin{enumerate}
\item[G.2] sampling a cause for each death to generate a new $\vec{Y} | \vec{F},\mathbf{S}$: 
\begin{equation}
\label{eqn:insilicoGibbsY}
 y_j=c_n | \vec{F},\vec{S}_j \sim \mbox{Multinomial}_N(1,\vec{L}_j) 
\end{equation}
where 
\begin{equation}
\label{insilicoGibbsL}
\ell_{j,n}  =  \frac{f_n \cdot  \prod_{k=1}^{K}{\left( \Pr(s_{j,k} | y_j=c_n)^{s_{j,k}} \cdot \left[1 - \Pr(s_{j,k} | y_j=c_n)\right]^{(1-s_{j,k})} \right)} }
{ \sum_{n'=1}^{N} f_{n'} \cdot \prod_{k=1}^{K}{\left( \Pr(s_{j,k} | y_j=c_{n'})^{s_{j,k}} \cdot \left[1 - \Pr(s_{j,k} | y_j=c_{n'})\right]^{(1-s_{j,k})} \right)}}
\end{equation}
\item[G.3] sampling a new $\vec{F} | \vec{Y}, \mathbf{S}$:
\begin{equation}
\label{eqn:insilicoGibbsF}
\vec{F} | \vec{Y}, \mathbf{S} \sim \mbox{Dirichlet}(\vec{M}+\vec{\alpha})
\end{equation}
\end{enumerate}
The resulting sample of $(\vec{F},\vec{Y})$ and the $\vec{L_j}$ that go with it form the output of the method.  These are distributions of CSMFs at the population level and probabilities of dying from each cause at the individual level. These distributions can be summarized as required to produce point values and measures of uncertainty in $\vec{F}$ and $\vec{L_j}$.  

Deaths often result from more than one cause.  InSilicoVA accommodates this possibility by producing a separate distribution of the probabilities of being assigned to each cause; that is $N$ distributions, one for each cause.  In contrast, InterVA reports one value for each cause, and those values sum to unity across causes for a single death. 

Finally, with a suitable $\vec{F}$, InSilicoVA can be used to assign causes (and their associated $\ell_{j,n}$) to a single death by repeatedly drawing causes using (\ref{eqn:insilicoGibbsY}). This requires no more information than InterVA to accomplish the same objective, and it produces uncertainty bounds around the probabilities of being assigned to each cause.

\section{Testing \& Comparing InSilicoVA and InterVA}\label{sec:methods}

To evaluate both InSilicoVA and InterVA we fit them to simulated and real data.  We have created R code that implements both methods.  The R code for InterVA matches the results produced by Peter Byass' implementation \citep{2013interVA}. 

\subsection{Simulation Study}\label{sec:simStudy}

Our simulated data are generated using this procedure:
\begin{enumerate}
\item Draw a set of $\Pr(s_k | c_n)$ so that they have the same distribution and range as those provided with Byass' InterVA software \citep{2013interVA}.
\item Draw a set of simulated deaths from a made up distribution of deaths by cause.
\item For each simulated death, assign a set of signs/symptoms by applying the conditional probabilities simulated in step 1.
\end{enumerate}
These simulated data have the same overall features as the data required for either InterVA or InSilicoVA, and we know both the real population distribution of deaths by cause and the true individual cause assignments. 

Our simulation study poses three questions:
\begin{enumerate}
\item {\bf Fair comparison of InSilicoVA and InterVA}.  To make this comparison we generate 100 simulated datasets and apply both methods to each dataset.  We summarize the results with individual-level and population-level error measures.  We refer to this as `fair' because we apply both methods in their simplest form to data that fulfill all the requirements of both methods.  Since the data are effectively `perfect' we expect both methods to perform well.
\item {\bf Influence of numeric values of $\Pr(s_k | c_n)$}. Given the structure of InterVA, we are concerned that the results of InterVA may be sensitive to the exact numerical values taken by the conditional probabilities supplied by physicians.  To test this we rescale the simulated $\Pr(s_k | c_n)$ so that their range is restricted to $[0.25-0.75]$.  We again simulate 100 datasets and apply both methods to each dataset and summarize the resulting errors for each method.
\item {\bf Reporting errors}. As we mentioned in the introduction, we are concerned about reporting errors for any algorithmic approach to VA cause assignment.  To investigate this we randomly recode our simulated signs/symptoms so that a small fraction are `wrong' - i.e. coded 0 when the sign/symptom exists (15\%) or 1 when there is no sign/symptom (10\%).  We do this for 100 simulated datasets and summarize errors resulting from application of both methods. 
\end{enumerate}

\subsection{Application to Real Data - Agincourt HDSS}\label{sec:aginStudy}

To investigate the behavior of InSilicoVA and InterVA on real data, we apply both methods to the VA data generated by the Agincourt health and demographic surveillance system (HDSS) in South Africa \citep{kahn2012profile} from roughly 1993 to the present.  The Agincourt site continuously monitors the population of 21 villages located in the Bushbuckridge District of Mpumalanga Province in northeast South Africa. This is a rural population living in what was during Apartheid a black `homeland'. The Agincourt HDSS was established in the early 1990s with the purpose of guiding the reorganization of South Africa's health system. Since then the goals of the HDSS have evolved and now it contributes to evaluation of national policy at population, household and individual levels. The population covered by the Agincourt site is approximately eighty-thousand.

For this test we us the physician-generated conditional probabilities $\mathbf{P}$ and initial guess of the CSMFs $\vec{F'}$ provided by Byass with the InterVA software \citep{2013interVA}.  

\subsection{Results}\label{sec:results}

\begin{figure}[H]
\begin{minipage}[b]{0.5\linewidth}
\centering
\includegraphics[width=1\textwidth]{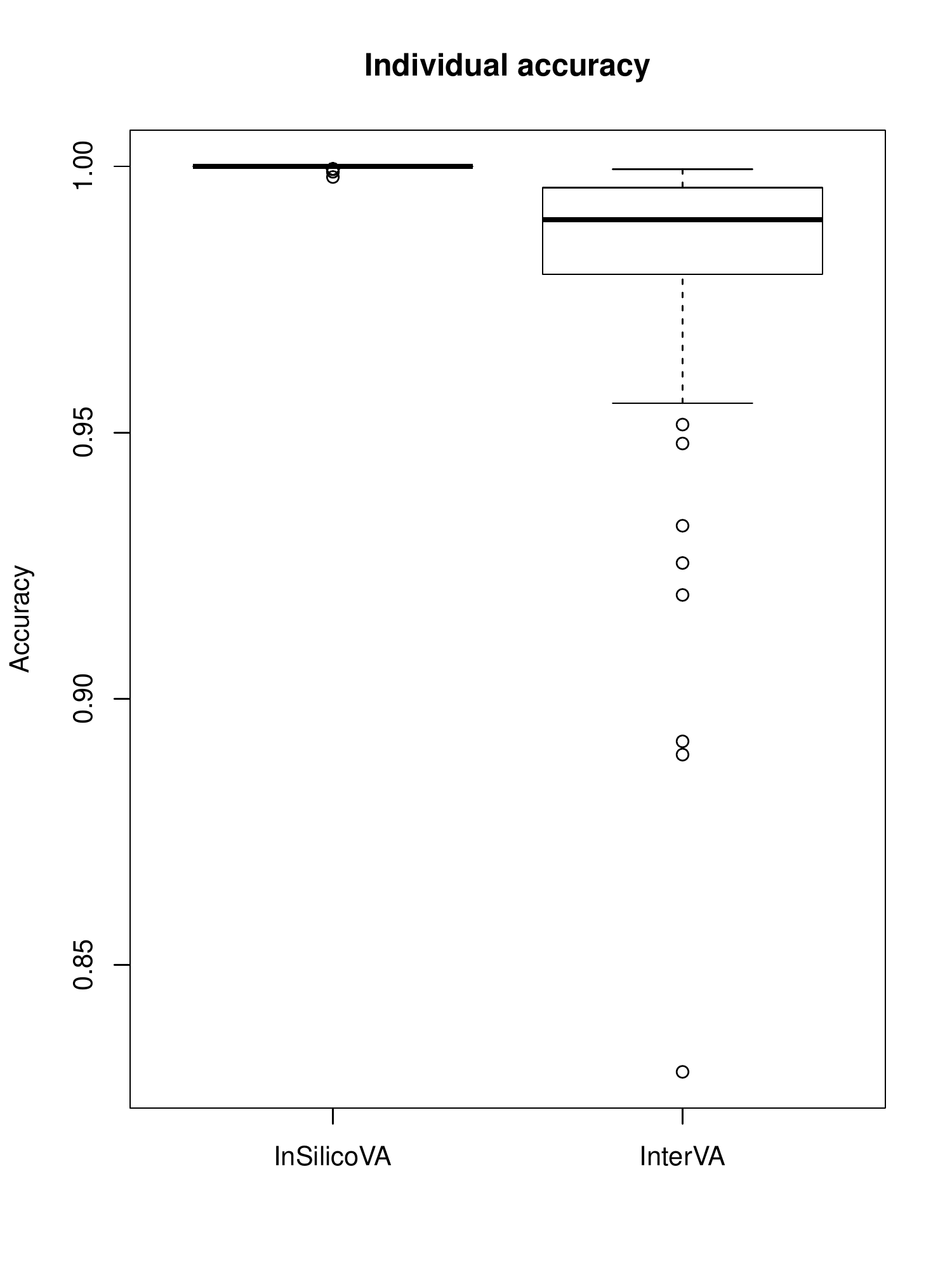}
{\bf A}
\end{minipage}
\hspace{0.5cm}
\begin{minipage}[b]{0.5\linewidth}
\centering
\includegraphics[width=1\textwidth]{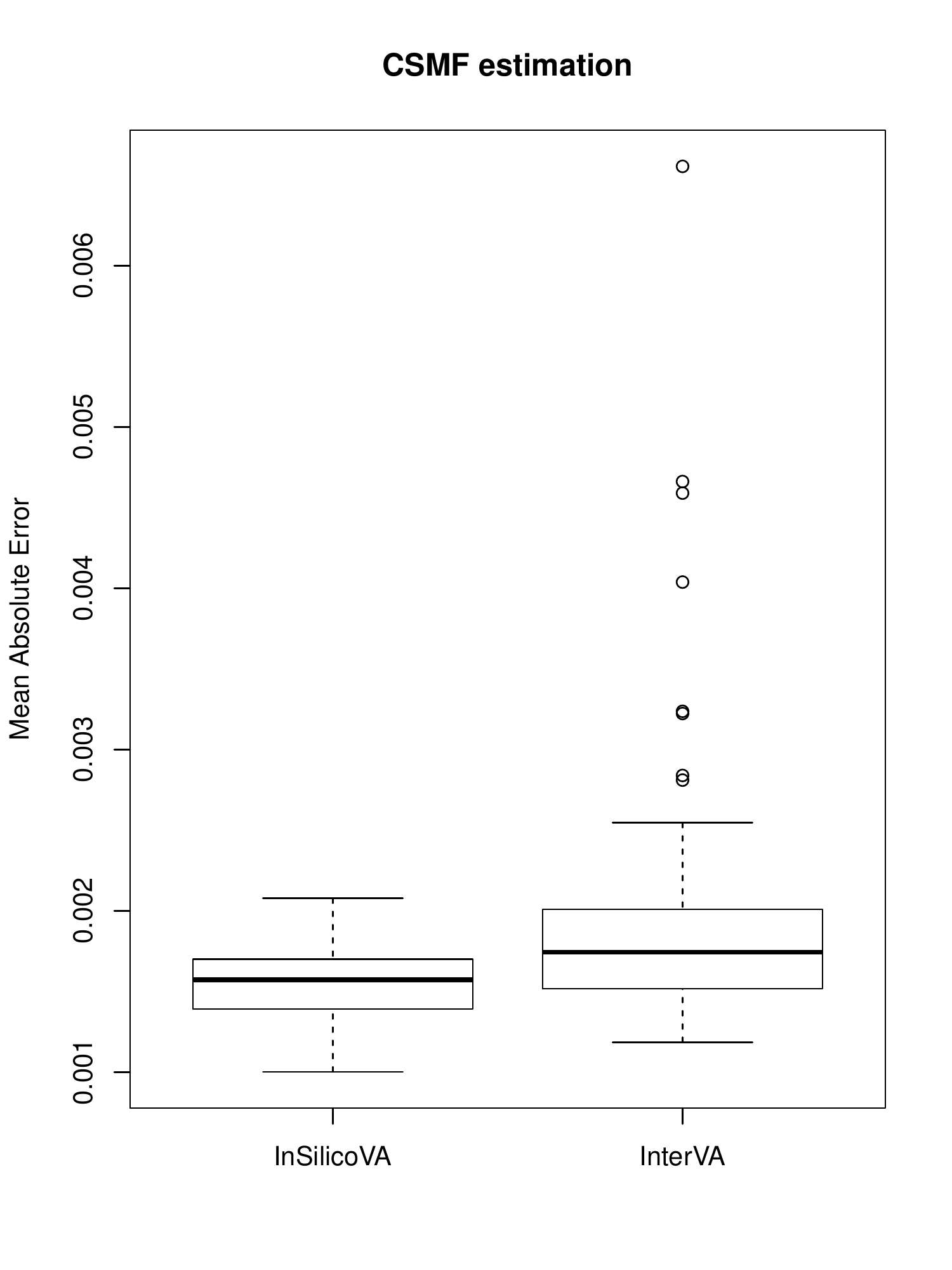}
{\bf B}
\end{minipage}
\caption{{\bf Simulation setup 1: `Fair Comparison'.} {\bf A}: InSilicoVA correctly assigns cause of death correctly effectively 100\% of the time.  InterVA is less accurate in assigning individual causes of death. {\bf B}: InSilicoVA's errors in identifying CSMFs are consistently very small.  InterVA's errors are also generally small, but the distribution has a long tail in the direction of large errors -- sometimes InterVA's errors are large.}
\label{fig:sim1}
\end{figure}

\begin{figure}[H]
\begin{minipage}[b]{0.5\linewidth}
\centering
\includegraphics[width=1\textwidth]{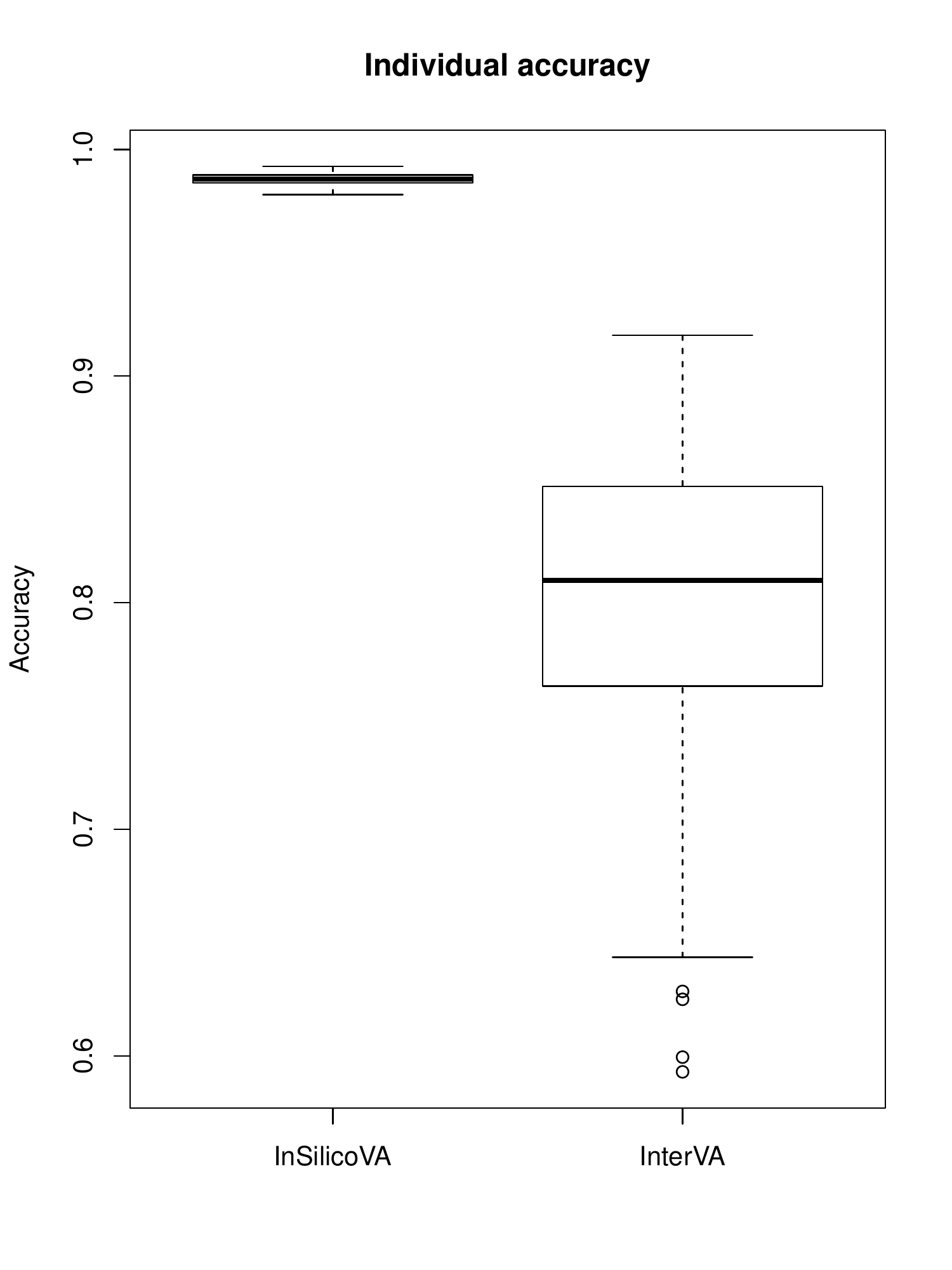}
\center{{\bf A}}
\end{minipage}
\hspace{0.5cm}
\begin{minipage}[b]{0.5\linewidth}
\centering
\includegraphics[width=1\textwidth]{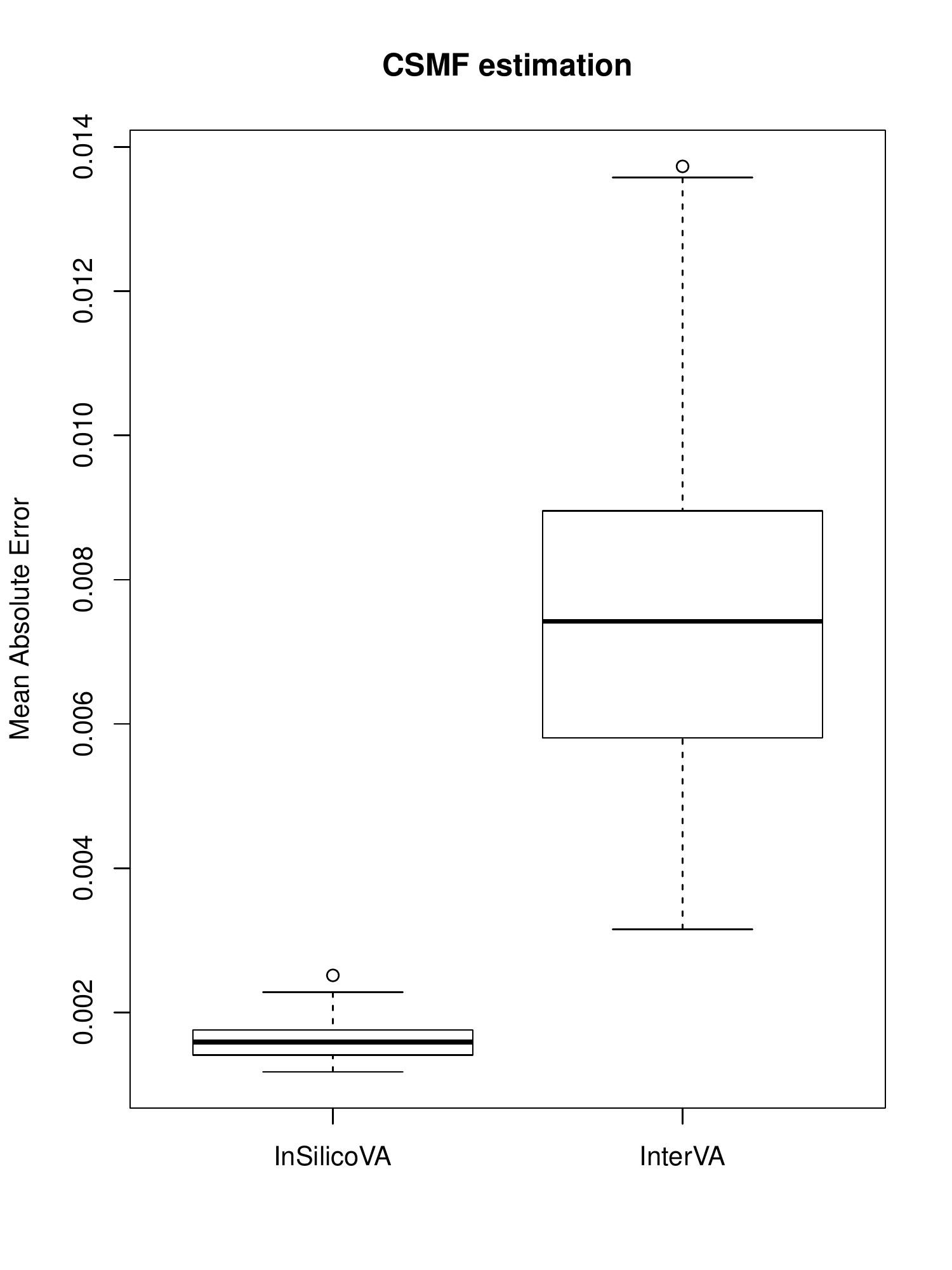}
\center{{\bf B}}
\end{minipage}
\caption{{\bf Simulation setup 2: `Conditional Probabilities in the range $[0.25-0.75]$'.} {\bf A}: InSilicoVA correctly assigns cause of death correctly effectively 100\% of the time.  InterVA correctly assigns cause of death correctly 80\% of the time with wide variation all the way down to as low as 60\% and never above about 90\%. {\bf B}: InSilicoVA's errors in identifying CSMFs are consistently very small.  InterVA's errors in identifying the CSMFs are larger and more variable.}
\label{fig:sim2}
\end{figure}

\begin{figure}[H]
\begin{minipage}[b]{0.5\linewidth}
\centering
\includegraphics[width=1\textwidth]{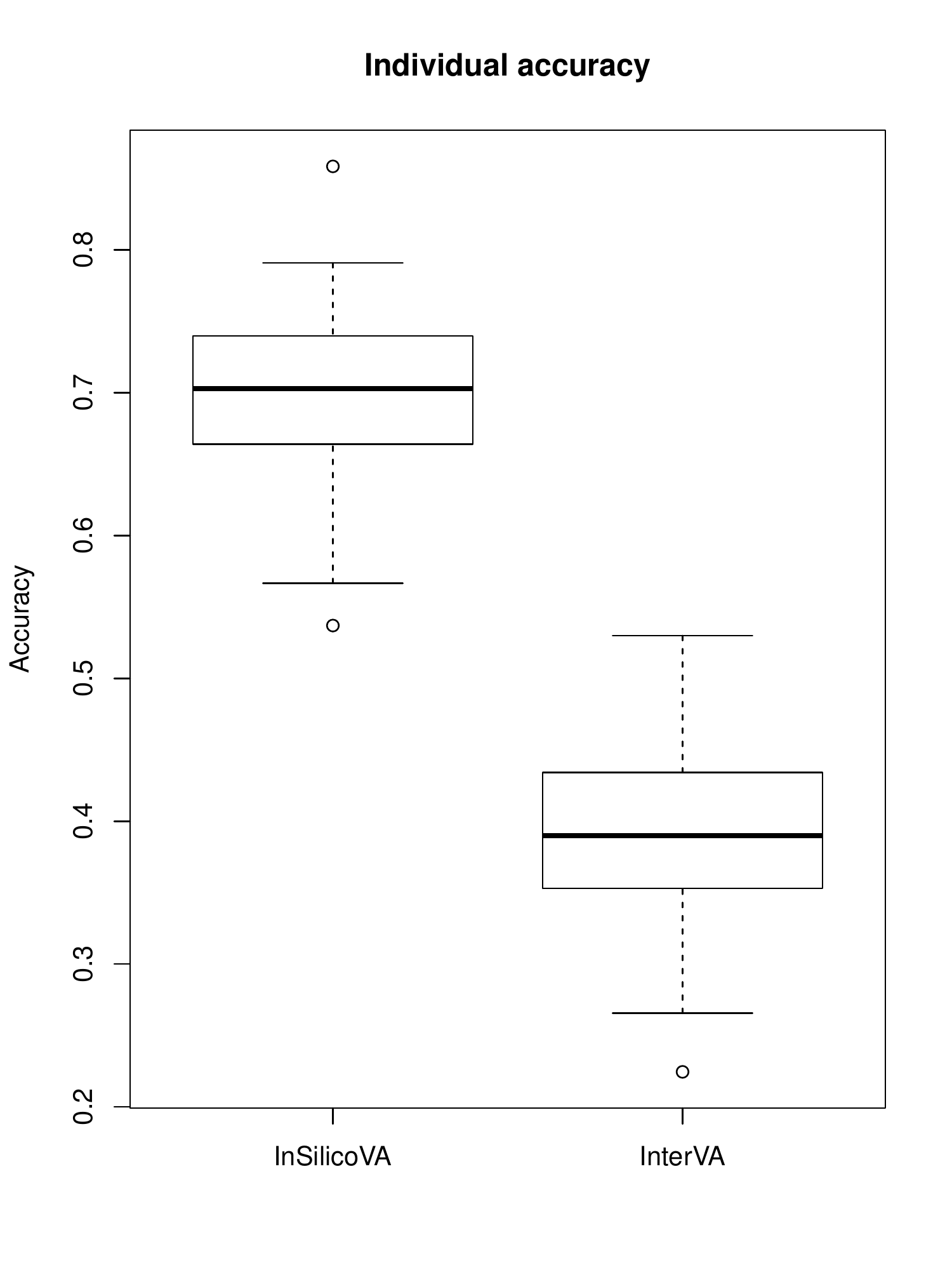}
\center{{\bf A}}
\end{minipage}
\hspace{0.5cm}
\begin{minipage}[b]{0.5\linewidth}
\centering
\includegraphics[width=1\textwidth]{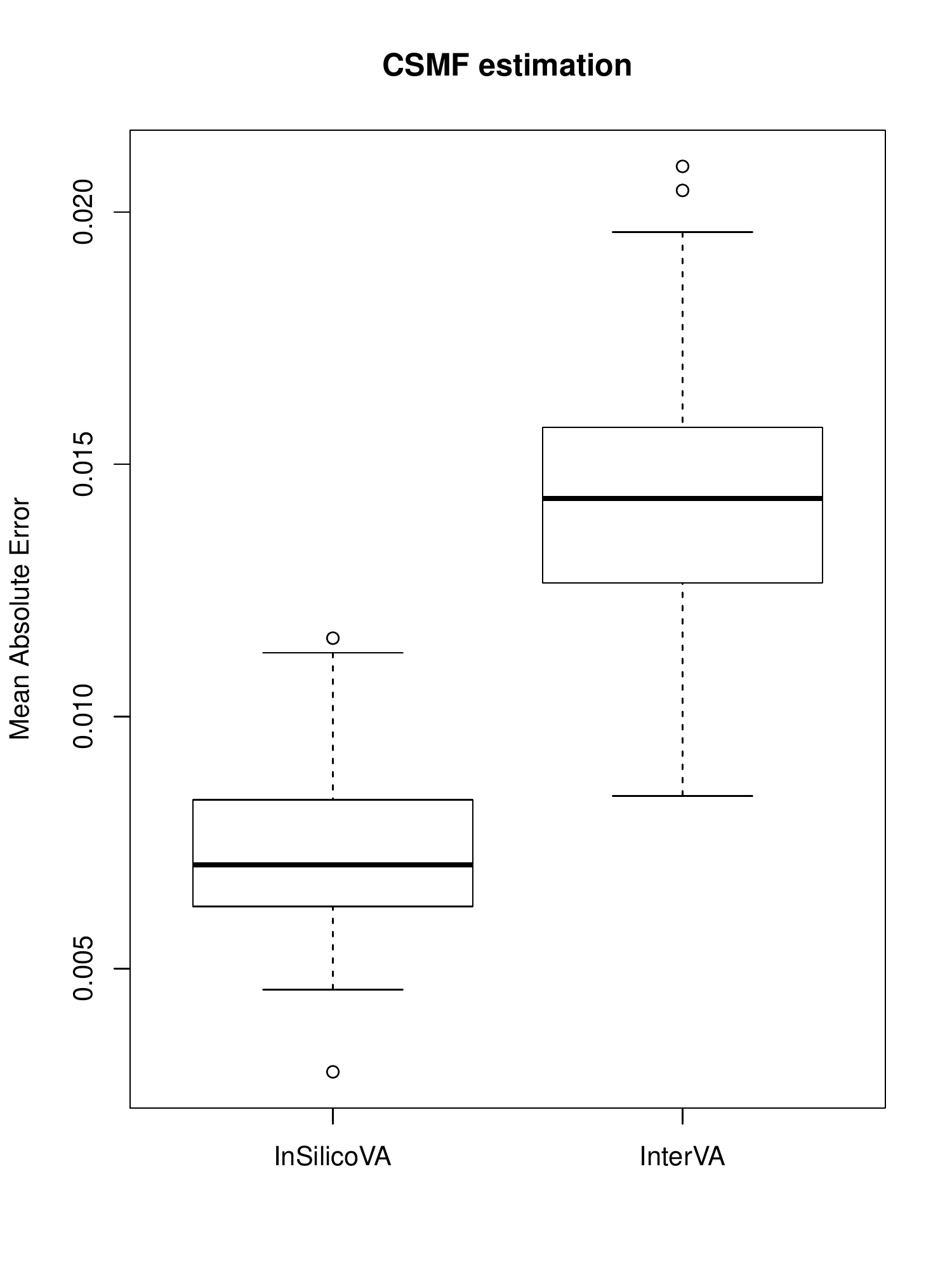}
\center{{\bf B}}
\end{minipage}
\caption{{\bf Simulation setup 3: `Reporting Errors'.} Both methods suffer, but InterVA suffers a lot more. {\bf A}: InSilicoVA correctly assigns cause of death correctly about 70\% of the time.  InterVA correctly assigns cause of death correctly about 40\% of the time. {\bf B}: InSilicoVA's errors in identifying CSMFs are still consistently  small.  InterVA's errors in identifying the CSMFs are larger.}
\label{fig:sim3}
\end{figure}

\begin{figure}[H]
\begin{minipage}[b]{0.5\linewidth}
\centering
\includegraphics[width=1\textwidth]{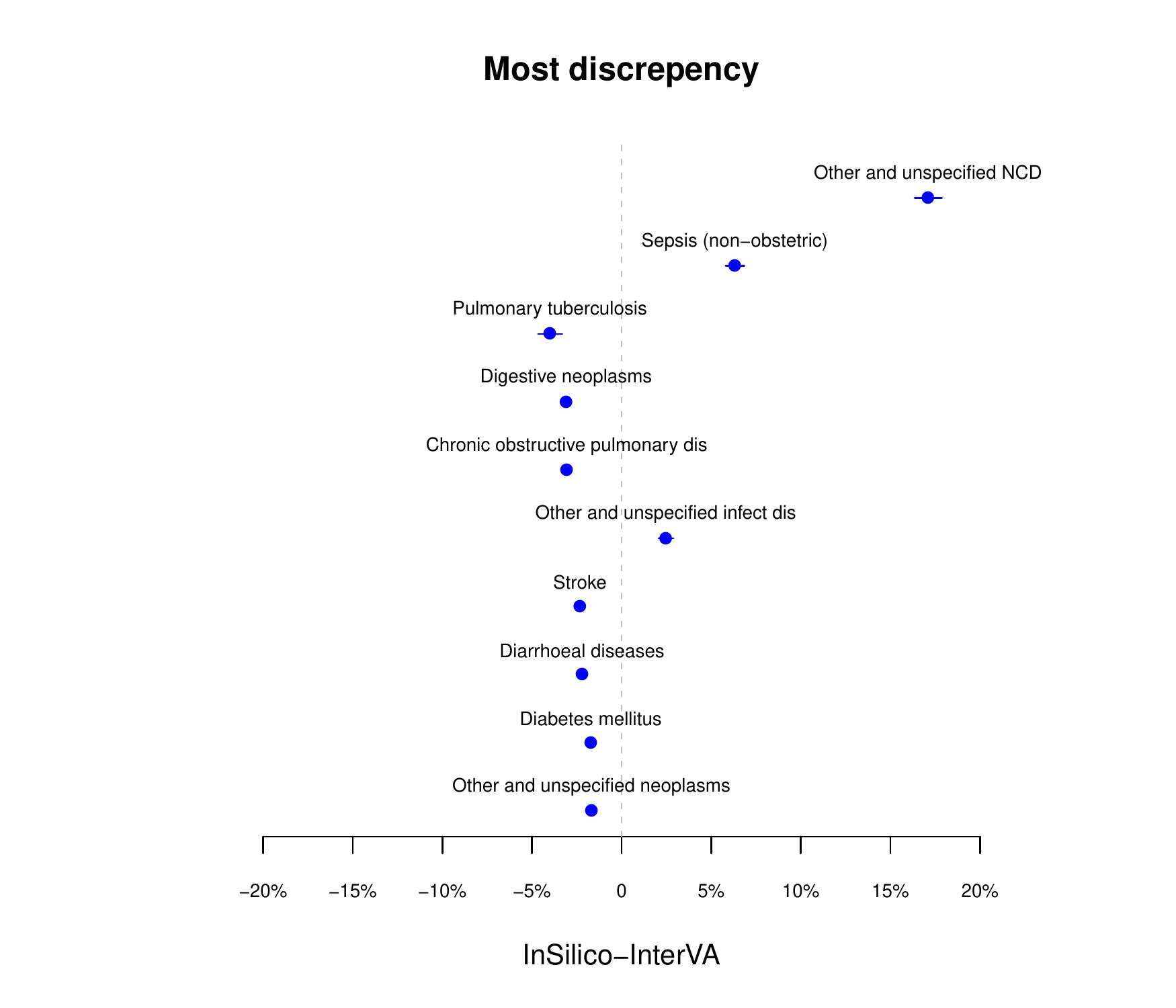}
{\bf A}
\end{minipage}
\hspace{0.5cm}
\begin{minipage}[b]{0.5\linewidth}
\centering
\includegraphics[width=1\textwidth]{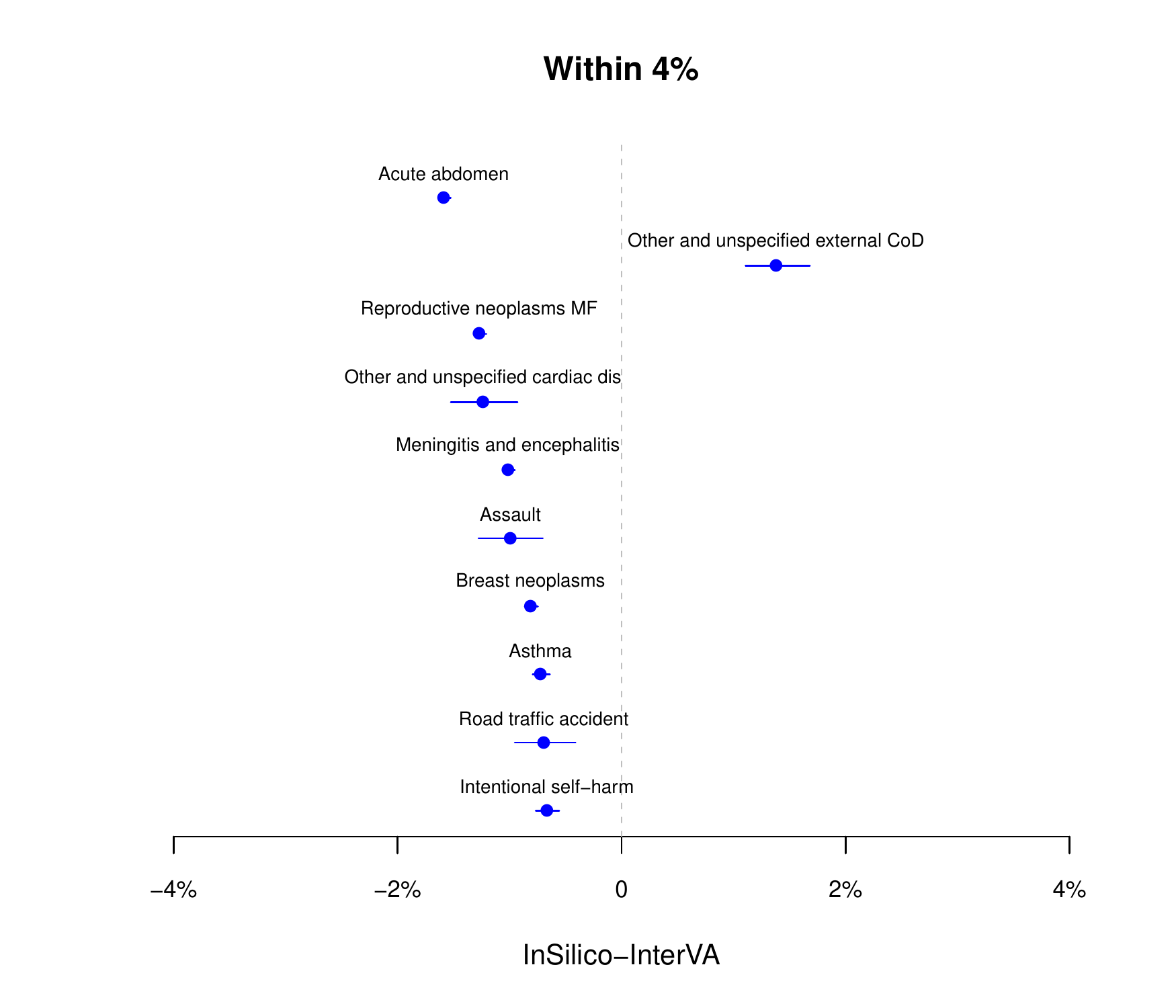}
{\bf B}
\end{minipage}
\begin{minipage}[b]{0.5\linewidth}
\vspace{1cm}
\centering
\includegraphics[width=1\textwidth]{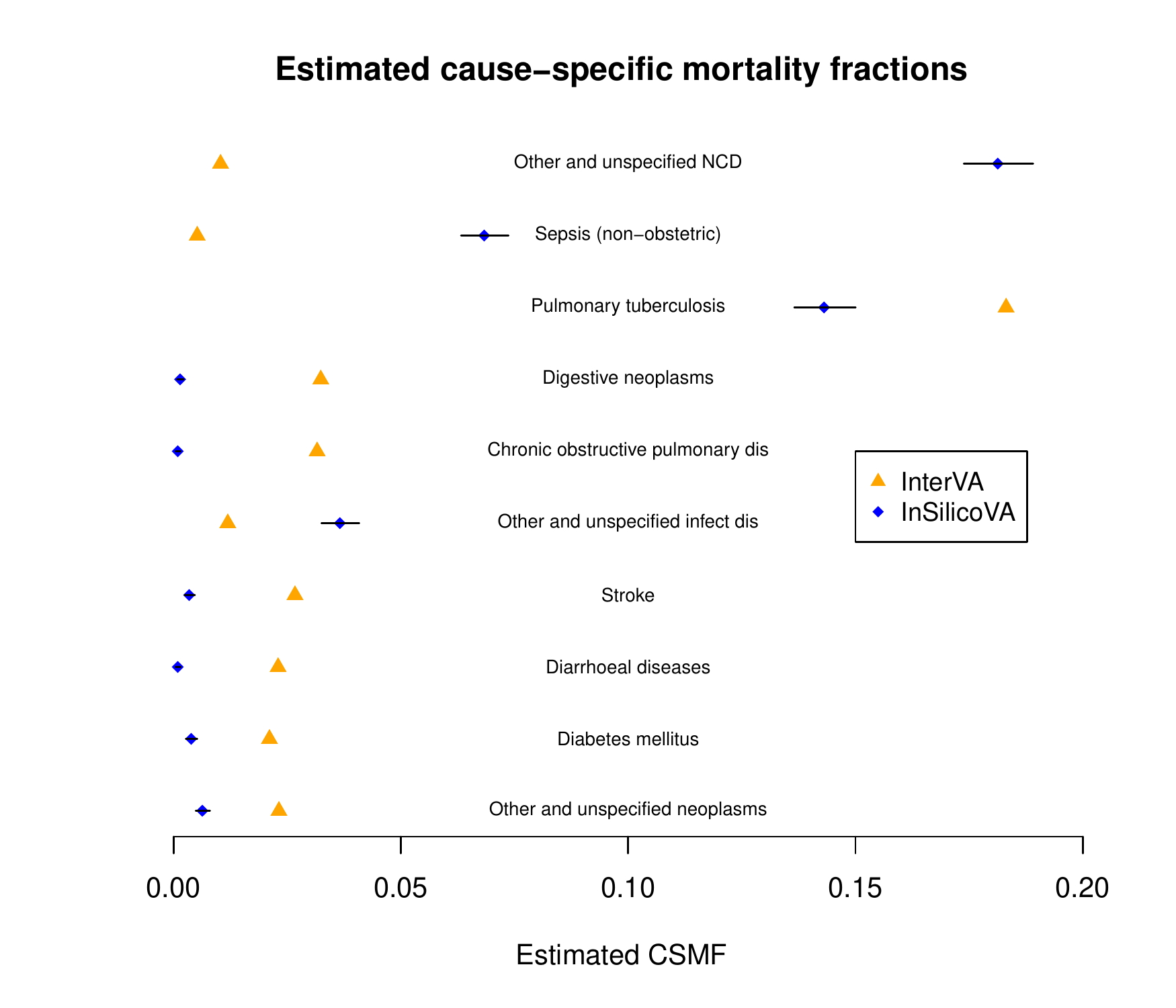}
{\bf C}
\end{minipage}
\caption{{\bf Agincourt Application.} {\bf A}: Differences in CSMFs (InSilicoVA -- InterVA) displaying the 10 specific causes that differ most.  InSilicoVA is less willing to make highly specific classifications and thus produces larger CSMFs associated with less-specific causes and smaller CSMFs associated with more specific causes. {\bf B}: Same as (A) but for the 10 largest differences in CSMFs that are still less than 4\%, which as a group includes HIV even though it is not in the top 10 that are plotted here. {\bf C}: The CSMFs produced by both models on their natural scale, the same 10 causes as in {\bf A} for which the differences are greatest. }
\label{fig:agin}
\end{figure}

The results of our simulation study are summarized graphically as a set of Figures \ref{fig:sim1} -- \ref{fig:sim3}.  The Agincourt results are presented in Figure \ref{fig:agin}.

\section{Discussion}\label{sec:discuss}

\subsection{Summary of Findings}

InSilicoVA begins to solve most of the critical problems with InterVA.  The results of applying both methods to simulated data indicate that InSilicoVA performs well under all circumstances except `reporting errors', but even in this situation InSilicoVA performs far better than InterVA.  InSilicoVA and InterVA both perform relatively well when the simulated data are perfect.  InSilicoVA's performance is not affected by changing the magnitudes and ranges of the conditional probability inputs, whereas InterVA's performance suffers dramatically.  With reporting errors both methods' performance is negatively impacted, but InterVA becomes effectively useless.  


Applied to one specific real dataset, both methods produce qualitatively similar results, but InSilicoVA is far more conservative and produces confidence bounds, whereas InterVA does not.  For Agincourt, Figure \ref{fig:agin}.A shows the causes with the largest difference between the InSilicoVA and InterVA estimates of the CSMF.  
InSilicoVA classifies a larger portion of deaths as due to causes labeled as `other.'  This indicates that these causes are related to either the communicable or non-communicable diseases, but there is not enough information to make a more specific classification.  This feature of InSilicoVA identifies cases that are difficult to classify using available data and may, for example, be good candidates for physician review.  

We view this behavior as a strength of InSilicoVA because it is consistent with the fundamental weakness of the VA approach, namely that both the information obtained from a VA interview and the expert knowledge and/or gold standard used to characterize the relationship between signs/symptoms and causes are inherently weak and incomplete, and consequently it is very difficult or impossible to make highly specific cause assignments using VA.  Given this, we do not want a method that is artificially precise, i.e. forces fine-tuned classification when there is insufficient information.  Hence we view InSilicoVA's behavior as reasonable, `honest' (in that it does not over interpret the data) and useful.  `Useful' in the sense that it identifies where our information is particularly weak and therefore where we need to apply more effort either to data or to interpretation, like addition physician reviews.

\subsection{Future Work}

We plan a variety of additional work on InSilicoVA:
\begin{enumerate}
\item Explore the possibility of replacing the Dirichlet distribution in (\ref{eqn:insilicoCsfmAssignDirichlet}), (\ref{eqn:insilicoDrawCsmf}) and (\ref{eqn:insilicoGibbsF}) with a mixture of Normals on the baseline logit transformed set of $f_n$'s.  This provides additional flexible parameters to allow each CSMF to have its own mean and variance.
\item Embed InSilicoVA in a spatio-temporal model that allows $\vec{F}$ to vary smoothly through space and time.  This would provide a parsimonious way of exploring spatio-temporal variation in the CSMFs while using the data as efficiently as possible.
\item Create the ability to add physician cause assignments to (\ref{eqn:insilicoIndividualProbCause}) and (\ref{insilicoGibbsL}) so that information in that form can be utilized when available.  The physician codes will require pre-processing to remove physician-specific bias in cause assignment, perhaps using a `rater reliability method' \citep[for example:][]{salter2012sentiment}.
\item {\bf Most importantly}, address the obviously invalid assumption that the signs/symptoms are independent given a specific cause.  This will require  modeling of the signs/symptoms and the physician-provided conditional probabilities so that important dependencies can be accommodated.  Further, this will require additional consultation with physicians and acquisition of new expert knowledge to characterize these dependencies.  All of this will require a generous grant and the collaboration of a large number of experts.  This will very likely greatly improve the performance and robustness of the method.
\item {\bf Critically}, re-elicit the conditional probabilities $\mathbf{P}$ from physicians so that they are logically well-behaved, i.e. fully consistent with one another and their complements.
\item Focus and sharpen VA questionnaire. Quantify the influence of each sign/symptom to: (1) potentially eliminate low-value signs/symptoms and thereby make the VA interview more efficient, and/or (2) suggest sign/symptom `types' that appear particularly useful, and potentially suggest augmenting VA interviews based on that information.
\item Explore new possibilities for refining the conditional probabilities $\mathbf{P}$ and potentially for entirely new models.  
\end{enumerate}


\newpage

\newpage
\newpage

\newpage

\newpage
\bibliographystyle{chicago}
\bibliography{VAbib}

\begin{thebibliography}{}

\bibitem[\protect\citeauthoryear{Byass}{Byass}{2013}]{2013interVA}
Byass, P. (2013).
\newblock Interva software.
\newblock {\em www.interva.org\/}.

\bibitem[\protect\citeauthoryear{Byass, Chandramohan, Clark, D'Ambruoso,
  Fottrell, Graham, Herbst, Hodgson, Hounton, Kahn, et~al.}{Byass
  et~al.}{2012}]{byass2012strengthening}
Byass, P., D.~Chandramohan, S.~J. Clark, L.~D'Ambruoso, E.~Fottrell, W.~J.
  Graham, A.~J. Herbst, A.~Hodgson, S.~Hounton, K.~Kahn, et~al. (2012).
\newblock Strengthening standardised interpretation of verbal autopsy data: the
  new interva-4 tool.
\newblock {\em Global health action\/}~{\em 5}.

\bibitem[\protect\citeauthoryear{Byass, Fottrell, Huong, Berhane, Corrah, Kahn,
  Muhe, et~al.}{Byass et~al.}{2006}]{byass2006refining}
Byass, P., E.~Fottrell, D.~L. Huong, Y.~Berhane, T.~Corrah, K.~Kahn, L.~Muhe,
  et~al. (2006).
\newblock Refining a probabilistic model for interpreting verbal autopsy data.
\newblock {\em Scandinavian journal of public health\/}~{\em 34\/}(1), 26--31.

\bibitem[\protect\citeauthoryear{Byass, Huong, and Van~Minh}{Byass
  et~al.}{2003}]{byass2003probabilistic}
Byass, P., D.~L. Huong, and H.~Van~Minh (2003).
\newblock A probabilistic approach to interpreting verbal autopsies:
  methodology and preliminary validation in vietnam.
\newblock {\em Scandinavian Journal of Public Health\/}~{\em 31\/}(62 suppl),
  32--37.

\bibitem[\protect\citeauthoryear{Flaxman, Vahdatpour, Green, James, Murray, and
  {Consortium Population Health Metrics Research}}{Flaxman et~al.}{2011}]{flax}
Flaxman, A.~D., A.~Vahdatpour, S.~Green, S.~L. James, C.~J. Murray, and
  {Consortium Population Health Metrics Research} (2011).
\newblock Random forests for verbal autopsy analysis: multisite validation
  study using clinical diagnostic gold standards.
\newblock {\em Popul Health Metr\/}~{\em 9\/}(29).

\bibitem[\protect\citeauthoryear{James, Flaxman, Murray, and {Consortium
  Population Health Metrics Research}}{James et~al.}{2011}]{james}
James, S.~L., A.~D. Flaxman, C.~J. Murray, and {Consortium Population Health
  Metrics Research} (2011).
\newblock Performance of the tariff method: validation of a simple additive
  algorithm for analysis of verbal autopsies.
\newblock {\em Popul Health Metr\/}~{\em 9\/}(31).

\bibitem[\protect\citeauthoryear{Kahn, Collinson, G{\'o}mez-Oliv{\'e}, Mokoena,
  Twine, Mee, Afolabi, Clark, Kabudula, Khosa, et~al.}{Kahn
  et~al.}{2012}]{kahn2012profile}
Kahn, K., M.~A. Collinson, F.~X. G{\'o}mez-Oliv{\'e}, O.~Mokoena, R.~Twine,
  P.~Mee, S.~A. Afolabi, B.~D. Clark, C.~W. Kabudula, A.~Khosa, et~al. (2012).
\newblock Profile: Agincourt health and socio-demographic surveillance system.
\newblock {\em International journal of epidemiology\/}~{\em 41\/}(4),
  988--1001.

\bibitem[\protect\citeauthoryear{King and Lu}{King and Lu}{2008}]{king2}
King, G. and Y.~Lu (2008).
\newblock Verbal autopsy methods with multiple causes of death.
\newblock {\em Statistical Science\/}~{\em 100\/}(469).

\bibitem[\protect\citeauthoryear{King, Lu, and Shibuya}{King
  et~al.}{2010}]{king1}
King, G., Y.~Lu, and K.~Shibuya (2010).
\newblock Designing verbal autopsy studies.
\newblock {\em Popul Health Metr\/}~{\em 8\/}(19).

\bibitem[\protect\citeauthoryear{Leitao, Chandramohan, Byass, Jakob,
  Bundhamcharoen, and Choprapowan}{Leitao et~al.}{2013}]{leitao2013whoStandard}
Leitao, J., D.~Chandramohan, P.~Byass, R.~Jakob, K.~Bundhamcharoen, and
  C.~Choprapowan (2013).
\newblock Revising the {WHO} verbal autopsy instrument to facilitate routine
  cause-of-death monitoring.
\newblock {\em Global Health Action\/}~{\em 6\/}(21518).

\bibitem[\protect\citeauthoryear{Maher, Biraro, Hosegood, Isingo, Lutalo,
  Mushati, Ngwira, Nyirenda, Todd, and Zaba}{Maher
  et~al.}{2010}]{maher2010translating}
Maher, D., S.~Biraro, V.~Hosegood, R.~Isingo, T.~Lutalo, P.~Mushati, B.~Ngwira,
  M.~Nyirenda, J.~Todd, and B.~Zaba (2010).
\newblock Translating global health research aims into action: the example of
  the alpha network*.
\newblock {\em Tropical Medicine \& International Health\/}~{\em 15\/}(3),
  321--328.

\bibitem[\protect\citeauthoryear{Murray, James, Birnbaum, Freeman, Lozano,
  Lopez, and {Consortium Population Health Metrics Research}}{Murray
  et~al.}{2011}]{murray}
Murray, C.~J., S.~L. James, J.~K. Birnbaum, M.~K. Freeman, R.~Lozano, A.~D.
  Lopez, and {Consortium Population Health Metrics Research} (2011).
\newblock Simplified symptom pattern method for verbal autopsy analysis:
  multisite validation study using clinical diagnostic gold standards.
\newblock {\em Popul Health Metr\/}~{\em 9\/}(30).

\bibitem[\protect\citeauthoryear{Salter-Townshend and Murphy}{Salter-Townshend
  and Murphy}{2012}]{salter2012sentiment}
Salter-Townshend, M. and T.~B. Murphy (2012).
\newblock Sentiment analysis of online media.
\newblock {\em Lausen, B., van del Poel, D. and Ultsch, A.(eds.). Algorithms
  from and for Nature and Life. Studies in Classification, Data Analysis, and
  Knowledge Organization\/}.

\bibitem[\protect\citeauthoryear{Sankoh and Byass}{Sankoh and
  Byass}{2012}]{Sankoh2012}
Sankoh, O. and P.~Byass (2012).
\newblock The indepth network: filling vital gaps in global epidemiology.
\newblock {\em International Journal of Epidemiology\/}~{\em 41\/}(3),
  579--588.

\end{thebibliography}

\end{document}